\documentclass[a4paper,11pt]{article}
\usepackage{pos}

\usepackage{braket}
\usepackage{bm,bbm}

\newcommand\beq{ \begin{eqnarray} }
\newcommand\eeq{ \end{eqnarray} }

\title{Three ways of calculating mass spectra for the 2-flavor Schwinger model in the Hamiltonian formalism}
\ShortTitle{Three ways of calculating mass spectra in the Hamiltonian formalism}

\author*[a,b]{Akira Matsumoto}
\author[a,b]{Etsuko Itou,}
\author[a]{Yuya Tanizaki}

\affiliation[a]{
Yukawa Institute for Theoretical Physics, Kyoto University,
Sakyo-ku, Kyoto 606-8502, Japan}

\affiliation[b]{
Interdisciplinary Theoretical and Mathematical Sciences Program (iTHEMS), RIKEN,\\
Wako 351-0198, Japan}

\emailAdd{itou(at)yukawa.kyoto-u.ac.jp}
\emailAdd{akira.matsumoto(at)yukawa.kyoto-u.ac.jp}
\emailAdd{yuya.tanizaki(at)yukawa.kyoto-u.ac.jp}

\abstract{We propose three independent methods to compute the hadron mass spectra of gauge theories in the Hamiltonian formalism.
The determination of hadron masses is one of the key issues in QCD,
which has been precisely calculated by the Monte Carlo method in the Lagrangian formalism.
We confirm that the mass of hadrons can be calculated by examining correlation functions,
the one-point function, or the dispersion relation in Hamiltonian formalism.
These methods are suitable for quantum computation and tensor network approaches.
The methods are demonstrated with the tensor network (DMRG) in the 2-flavor Schwinger model,
which shares important properties with QCD.
We show that the numerical results are consistent with each other and with the analytic prediction of the bosonization technique.
We also discuss their efficiency and potential applications to other models.
This talk is based on the paper \cite{Itou:2023img}.}

\FullConference{%
  The 40th International Symposium on Lattice Field Theory (Lattice2023)\\
  31 July - 4 August, 2023 \\
  Fermilab, Batavia, Illinois, USA
}

\begin{document}

%%% preprint number
\begin{flushright}
YITP-23-154, RIKEN-iTHEMS-Report-24
\end{flushright}

\maketitle

\section{Introduction}
\label{sec:intro}

Numerical simulation of quantum field theories in the Hamiltonian formalism would be 
a promising approach thanks to recent developments in quantum computing and tensor network methods.
These methods are expected to be complementary frameworks to Monte Carlo simulations in the Lagrangian formalism 
since they are free from the sign problem originating from the importance sampling.
Taking advantage of the numerical methods in the Hamiltonian formalism, 
we develop three distinct ways to compute the mass spectrum 
of gauge theories; the correlation-function scheme, 
the one-point-function scheme, and the dispersion-relation scheme.
We summarize the properties of these methods based on the results in Ref.~\cite{Itou:2023img}, 
where the methods are demonstrated by applying them to the 2-flavor massive Schwinger model 
using the density-matrix renormalization group (DMRG) 
\cite{White:1992zz,White:1993zza,Schollw_ck_2005,Schollw_ck_2011}.

In the 2-flavor Schwinger model, there are three composite particles (mesons), 
pions $\pi_a$, sigma meson $\sigma$, and eta meson $\eta$.
The corresponding operators and quantum numbers are given by, 
\begin{align}
\pi_a & =-i \bar{\psi}\gamma^5 \tau_a \psi \ \left(J^{PG}=1^{-+}\right), &
\sigma & =\bar{\psi}\psi \ \left(J^{PG}=0^{++}\right), &
\eta & =-i \bar{\psi}\gamma^5\psi \ \left(J^{PG}=0^{--}\right),
\label{eq:meson_op}
\end{align}
where $J$, $P$, and $G$ represent isospin, parity, and $G$-parity, respectively.
We compute the mass spectrum to test the validity of the three methods 
and find that the results mostly agree with each other.
Our numerical results indicate that the sigma meson is stable unlike QCD 
since its mass is lighter than twice the pion mass.
This result is consistent with the analytic calculation 
using the WKB approximation of the bosonized model \cite{Coleman:1976uz}, 
where the ratio of $\pi$ and $\sigma$ mass is given by $M_\sigma/M_\pi=\sqrt{3}$.
Indeed, our computations give roughly consistent results of the mass ratio.

\section{Hamiltonian and calculation method}

Let us consider the Hamiltonian of the $N_f=2$ Schwinger model on the lattice 
with the staggered fermion \cite{Kogut:1974ag,Susskind:1976jm}.
In the Hamiltonian formalism, the physical Hilbert space is constrained by the Gauss law condition.
We choose the open boundary condition so that the lattice Gauss law equation can be solved explicitly.
Then the electric field operator is replaced by the integration of the charge density, 
which consists of fermion bilinear operators.
Furthermore, we can remove all the degrees of freedom of link variables from the Hamiltonian 
since they can be absorbed into the U(1) phase of the fermions by gauge fixing.
Thus, the Hamiltonian is written only by the fermionic operators.
Finally, we map the lattice Hamiltonian to the spin Hamiltonian using the Jordan-Wigner transformation, 
which is convenient for applying tensor network methods or quantum computations.
As a result, we obtain the spin Hamiltonian with a finite-dimensional Hilbert space, 
\begin{align}
H= & \frac{g^2a}{8}\sum_{n=0}^{N-2}
\left[\sum_{f=1}^{N_{f}}\sum_{k=0}^{n}\sigma_{f,k}^{z}
+N_{f}\frac{(-1)^{n}+1}{2}+\frac{\theta}{\pi}\right]^2 \nonumber \\
 & -\frac{i}{2a}\sum_{n=0}^{N-2}\left(
\sigma_{1,n}^{+}\sigma_{2,n}^{z}\sigma_{1,n+1}^{-}
%-\sigma_{1,n}^{-}\sigma_{2,n}^{z}\sigma_{1,n+1}^{+}
+\sigma_{2,n}^{+}\sigma_{1,n+1}^{z}\sigma_{2,n+1}^{-}
%-\sigma_{2,n}^{-}\sigma_{1,n+1}^{z}\sigma_{2,n+1}^{+}
-\mathrm{h.c.}\right)
+\frac{m_{\mathrm{lat}}}{2}\sum_{f=1}^{N_f}\sum_{n=0}^{N-1}(-1)^{n}\sigma_{f,n}^z.
\end{align}
Here the mass $m$ of the continuum theory is replaced by 
$m_{\mathrm{lat}}:=m-N_{f}g^{2}a/8$ in the lattice Hamiltonian, 
following the proposal \cite{Dempsey:2022nys} for eliminating $O(a)$ correction.

We use the DMRG method to obtain the ground state of the spin Hamiltonian.
The DMRG is a variational algorithm using the matrix product state (MPS) as an ansatz, 
where the MPS is updated to decrease the energy $E=\Braket{\Psi|H|\Psi}$.
In addition, the low-rank approximation by the singular value decomposition (SVD) is performed.
We introduce a cutoff parameter $\varepsilon$ to determine the bound dimension 
such that the truncation error by SVD is smaller than $\varepsilon$.
The approximation by MPS is improved by setting $\varepsilon$ smaller, 
whereas it requires more computational cost due to the growth of the bond dimension.
It is also possible to obtain the low-energy excited states using DMRG.
For this purpose, we change the Hamiltonian in DMRG for the $\ell$-th excited state as 
\begin{equation}
H_{\ell}=H+W\sum_{\ell^{\prime}=0}^{\ell-1}\ket{\Psi_{\ell^{\prime}}}\bra{\Psi_{\ell^{\prime}}},  
\end{equation}
where $W>0$ is a weight to impose the orthogonality. 
We can obtain the eigenstates of the Hamiltonian up to any level step by step. 
We used the C++ library of ITensor \cite{itensor} to perform the tensor network calculation.

\section{Simulation results}
\label{sec:result}

In this section, we show the numerical results of the mass spectrum 
of the 2-flavor Schwinger model at $\theta=0$ obtained by the three methods.
From now on, we set the gauge coupling $g=1$ to measure the energy scale in this unit 
since $g$ has mass dimension $1$. The fermion mass is fixed to $m=0.1$.

\subsection{Correlation-function scheme}
\label{subsec:result_CF}

\begin{figure}[tb]
\centering
\includegraphics[scale=0.4]{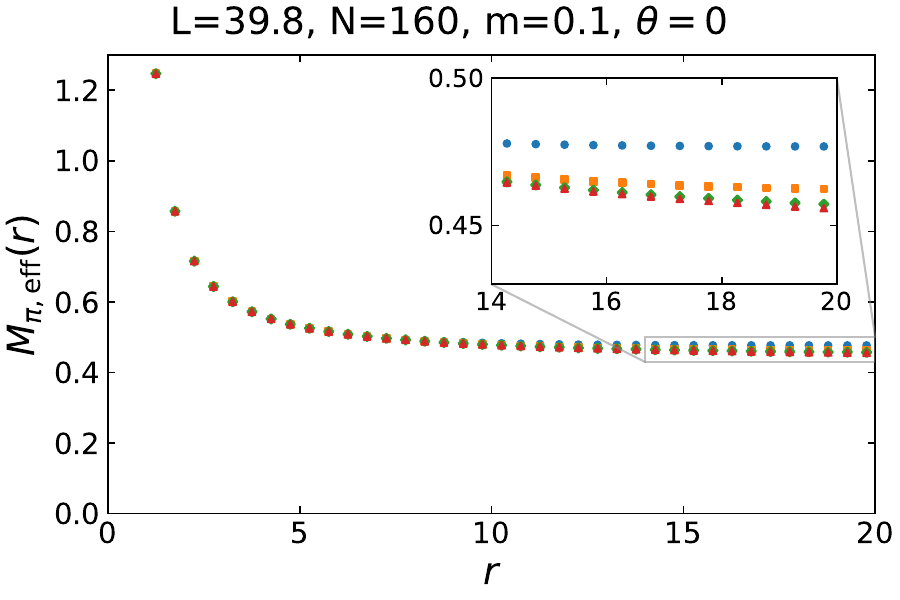}
\quad
\includegraphics[scale=0.4]{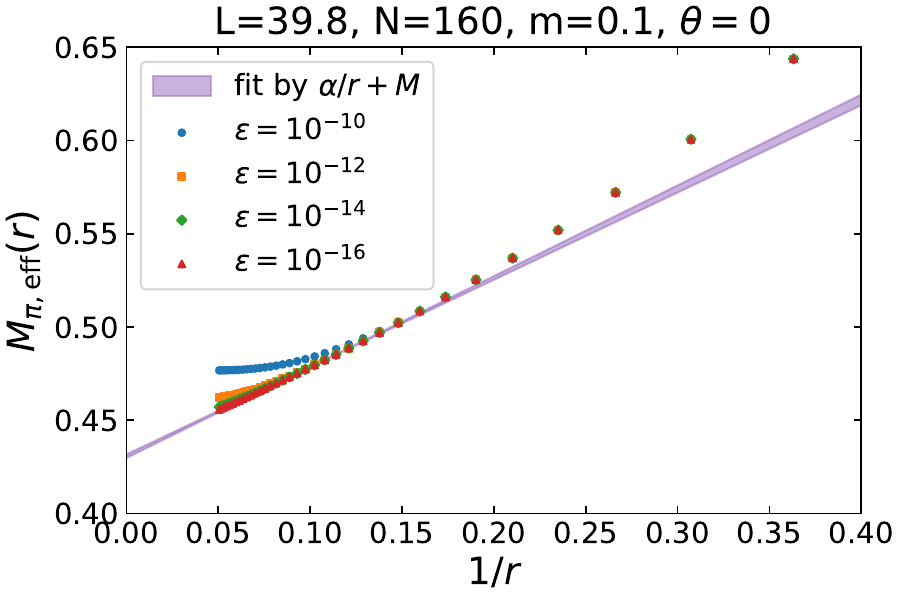}
\vspace{-0.4em}
\caption{\label{fig:fitMeff_pi}
(Left) The 3-point average of the pion effective mass $M_{\pi,\mathrm{eff}}(r)$ is 
plotted against the distance $r$ for different cutoff $\varepsilon$.
The number of lattice sites is $N=160$ 
and the lattice spacing $a$ is determined so that $L=a(N-1)=39.8$.
(Right) The effective mass $M_{\pi,\mathrm{eff}}(r)$ is plotted against $1/r$.
The data points for $\varepsilon=10^{-16}$ are fitted by $\alpha/r+M$ 
inside the region $0.075\protect\leq1/r\protect\leq0.15$.
The shaded band depicts the fitting result with the systematic error.}
\end{figure}

The mass spectrum is obtained from the spatial correlation function 
as we do in the conventional Euclidean lattice gauge theory.
First, we focus on the pion.
We measure the spatial correlation function for the ground state, 
$C_{\pi}(r)=\Braket{\pi(x)\pi(y)}$, by changing the distance $r=|x-y|$.
To investigate its exponential behavior, 
we compute the so-called effective mass $M_{\pi,\mathrm{eff}}(r)$.
The 3-point average of $M_{\pi,\mathrm{eff}}(r)$ is shown in the left panel of 
Fig.~\ref{fig:fitMeff_pi} for the various values of the cutoff parameter $\varepsilon$.

One usually regards the plateau value of $M_{\pi,\mathrm{eff}}(r)$ as the pion mass 
in the Monte Carlo study on the Euclidean lattice.
Although the result with the largest cutoff $\varepsilon=10^{-10}$ is 
almost flat for $r\gtrsim 10$, this plateau behavior is an artifact of the low-rank approximation.
Here the asymptotic form of the spatial correlator should be the Yukawa type, 
$C_{\pi}(r)\sim e^{-M_{\pi} r}/r^{\alpha}$, not purely exponential.
Thus, the corresponding effective mass is given by $M_{\pi,\mathrm{eff}}(r)\sim \alpha/r+M_\pi$.
To see this behavior, we plot $M_{\pi,\mathrm{eff}}(r)$ against $1/r$ 
in the right panel of Fig.~\ref{fig:fitMeff_pi}.
We find that the asymptotic behavior of the effective mass is sensitive to $\varepsilon$.
In fact, we can see the $1/r$-behavior of $M_{\pi,\mathrm{eff}}(r)$ 
for sufficiently small $\varepsilon$, namely the sufficiently large bond dimension.
The mass $M_\pi$ should be estimated by the linear extrapolation to $1/r\rightarrow 0$.
We fitted the data points for $\varepsilon=10^{-16}$ by $\alpha/r+M_\pi$, 
and obtained $M_\pi=0.431(1)$ and $\alpha=0.477(9)$ with the systematic error 
from the uncertainty of the fitting range.

Similarly, we can obtain the mass of the iso-singlet mesons.
We compute the effective mass of the sigma and eta mesons, 
$M_{\sigma,\mathrm{eff}}(r)$ and $M_{\eta,\mathrm{eff}}(r)$, 
and confirm that they behave as $\propto 1/r$ for smaller $\varepsilon$.
The results for $\varepsilon=10^{-16}$ are fitted by $\alpha/r+M$.
Then we obtained $M_\sigma = 0.722(6)$ with $\alpha=0.83(5)$ for sigma meson 
and $M_\eta=0.899(2)$ with $\alpha=0.51(2)$ for eta meson 
in the same way as for the pion.

\subsection{One-point-function scheme}
\label{subsec:result_1pt}

The second method to compute the meson mass does not 
rely on the two-point correlation functions.
Instead, we use the open boundary as a source of excitation 
from the thermodynamic ground state.
In the gapped system, the boundary effect decays exponentially 
with the distance $x$ from the boundary.
Thus, the one-point function of a local operator should behave as 
$\Braket{\mathcal{O}(x)} \sim e^{-Mx}$.
Here $M$ corresponds to the mass of the lightest particle 
of the same quantum number as $\mathcal{O}(x)$.

\begin{figure}[tb]
\centering
\includegraphics[scale=0.4]{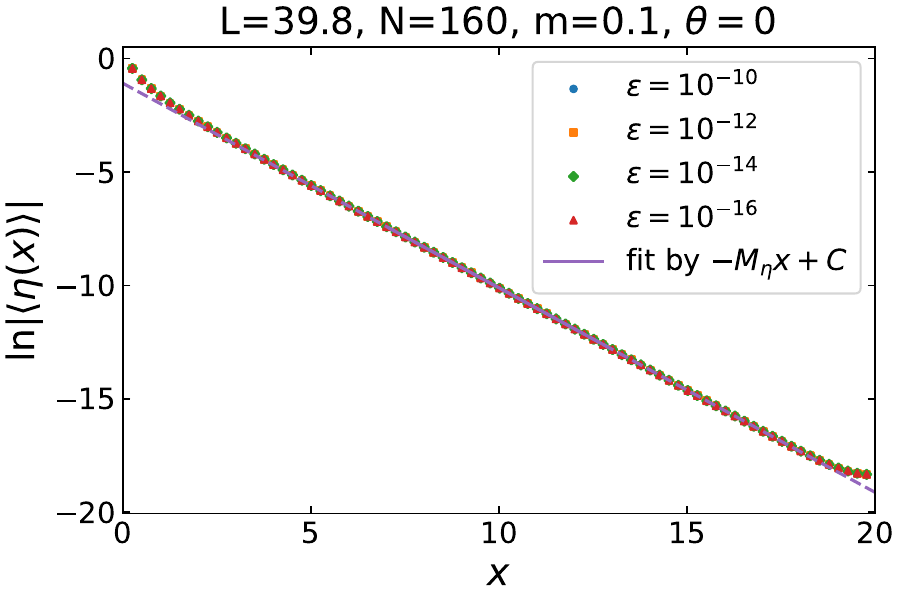}
\quad
\includegraphics[scale=0.4]{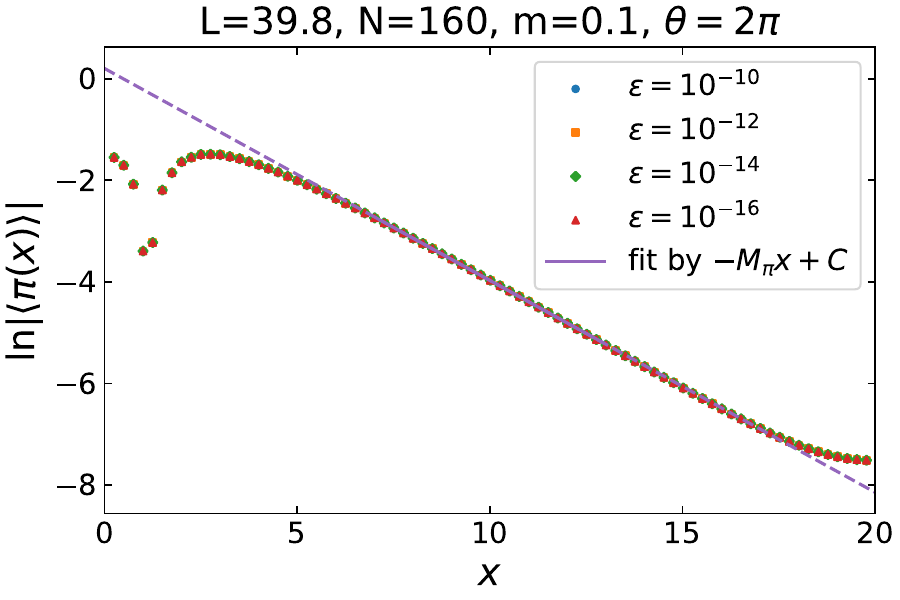}
\vspace{-0.4em}
\caption{\label{fig:1pt_eta}
The one-point functions of 
the eta meson $\ln|\Braket{\eta(x)}|$ at $\theta = 0$ (left) 
and the pion $\ln|\Braket{\pi(x)}|$ at $\theta = 2\pi$ (right) 
are plotted against the distance $x$ from the boundary for different cutoff $\varepsilon$.
The number of lattice sites is $N=160$, 
and the lattice spacing $a$ is determined so that $L=a(N-1)=39.8$.
The fitting results $-M_{\eta}x+C$ and $-M_{\pi}x+C$
for $\varepsilon=10^{-16}$ are depicted by the solid lines 
inside the fitting range and by the broken lines outside.}
\end{figure}

Let us first focus on the iso-singlet particles.
Since the $G$-parity is not spontaneously broken, 
the one-point functions of $\sigma$ and $\eta$ should be 
 $\Braket{\sigma(x)}=\mathrm{const.}$ and $\Braket{\eta(x)}=0$ in the thermodynamic limit.
However, the open boundary violates the $G$-parity, 
which is encoded as the one-unit lattice translation in the staggered fermion formalism.
Therefore, the boundary state is a source of the singlet mesons 
and gives nontrivial contributions to their one-point functions.
For example, the left panel of Fig.~\ref{fig:1pt_eta} shows 
the numerical result of $\Braket{\eta(x)}$ evaluated for the ground state 
with various cutoff parameters $\varepsilon$.
The one-point functions decay exponentially as expected.
Thus, we compute $\ln|\Braket{\sigma(x)-\sigma(L/2)}|$ and $\ln|\Braket{\eta(x)}|$ 
and fit them by $-M_{\sigma}x+C$ and $-M_{\eta}x+C$, respectively, 
in the range $7 \leq x \leq 13$.\footnote{
For the sigma meson, we subtract the value $\Braket{\sigma(L/2)}$ at the center $x=L/2$ 
of the lattice from $\Braket{\sigma(x)}$ to remove the constant shift.}
Then we obtain $M_{\sigma}=0.761(2)$ with $C=-2.71(2)$ for the sigma meson 
and $M_{\eta}=0.9014(1)$ with $C=-1.096(1)$ for the eta meson 
with the smallest cutoff $\varepsilon=10^{-16}$.
We find that the results with the other values of $\varepsilon$ are 
consistent within the fitting error. 
Thus, the cutoff dependence is negligible, unlike the case of the correlation functions.

Next, we turn to the analysis of the pion.
Since the boundary state at $\theta=0$ is invariant under the isospin rotation, 
it does not produce any single pion state.
As a result, we have $\Braket{\pi(x)}=0$ everywhere.
To induce the nontrivial $\Braket{\pi(x)}$, we need 
a boundary state that transforms nontrivially under the isospin rotation.
At $\theta=2\pi$, the ground state has a such property 
while the bulk property remains the same as $\theta=0$.
The ground state at $\theta=2\pi$ can be regarded as a nontrivial SPT state 
protected by the isospin symmetry.
Thus, there are boundary states with the isospin $1/2$, 
which can be a source of the pion, so that $\Braket{\pi(x)}\neq 0$. 
The one-point function $\Braket{\pi(x)}$ evaluated at $\theta=2\pi$ 
are shown in the right panel of Fig.~\ref{fig:1pt_eta}.
We confirm the result decays exponentially 
and fit the data points of $\ln|\Braket{\pi(x)}|$ 
by $-M_{\pi}x+C$ in the range $7\leq x\leq 13$.
The fitting results, $M_{\pi}=0.4175(9)$ and $C=0.203(9)$, 
are independent of $\varepsilon$ up to the fitting error as before.

\subsection{Dispersion-relation scheme}
\label{subsec:result_ex}

The third method to compute the spectrum is based on a different idea 
from the previous ones and is specific to Hamiltonian formalism.
We compute the energy gaps $\Delta E$ and momentum square $K^2$ 
of the various excited states by generating them by DMRG.
Then we measure their quantum numbers to identify the type of meson.
The mass spectrum can be obtained from the dispersion relation 
$\Delta E\simeq\sqrt{K^2+M^2}$.

\begin{figure}[tb]
\centering
\includegraphics[scale=0.4]{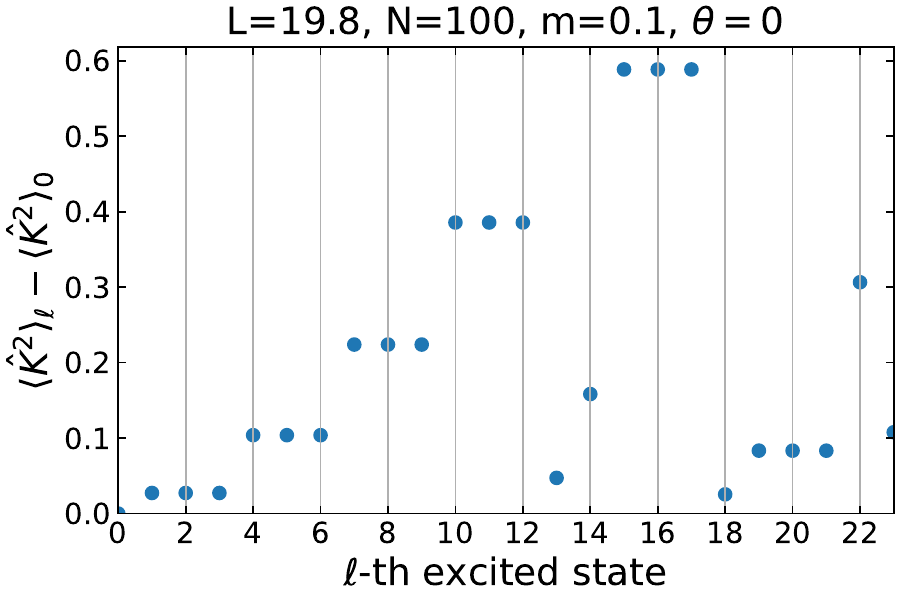}
\quad
\includegraphics[scale=0.4]{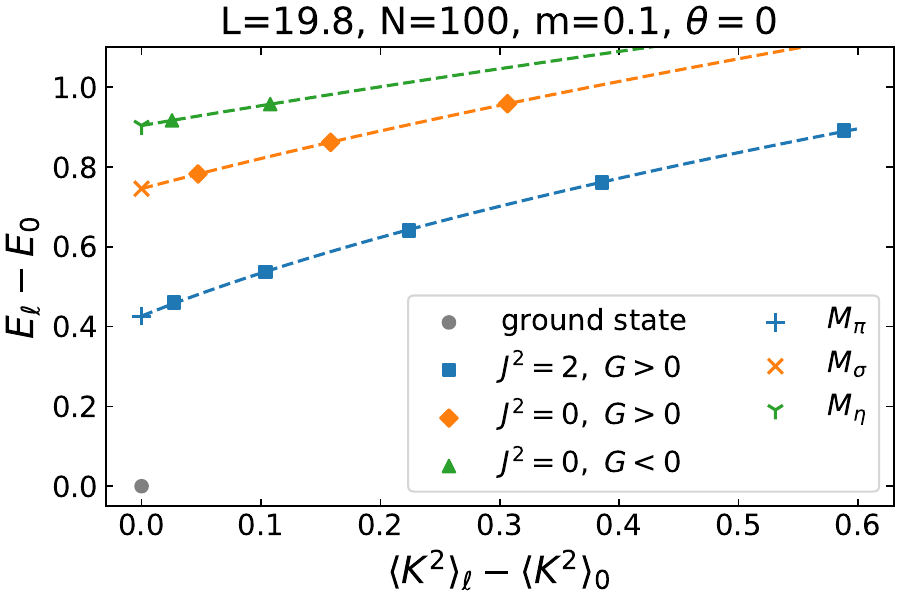}
\vspace{-0.4em}
\caption{\label{fig:E_and_K2}
(Left) The square of total momentum 
$\Delta K_{\ell}^2=\Braket{K^2}_{\ell}-\Braket{K^{2}}_{0}$ 
is plotted against the level of the excited state $\ell$ 
after subtracting the result of the ground state.
(Right) The energy gap $\Delta E_{\ell}$ is plotted against $\Delta K_{\ell}^{2}$.
The state with the same isospin and $G$-parity is represented by the same symbol.
The fitting results of the dispersion relations are depicted by the broken lines.}
\end{figure}

We generated the MPS up to the 23rd excited state at $\theta=0$ 
and measured the square of total momentum $K^2$.
The result is shown in the left panel of Fig.~\ref{fig:E_and_K2}.
From the energy gap and the total momentum, we find triply degenerated states, 
which seem to be the pion states.
There are also some singlet states, which will be the eta or sigma mesons.

Practically, we identify each state by measuring the expectation values of the isospin operators 
$\bm{J}^2$ and $J_z$, the parity $P$, and the $G$-parity $G=Ce^{i\pi J_y}$.
We find that the expectation values of the isospin operators are 
$\bm{J}^2=2$ with $J_z=0,\pm1$ for the triplets 
and $\bm{J}^2=0$ with $J_z=0$ for the singlets precisely.
On the other hand, we have $|P|,|G|\neq1$ due to the boundary effect.
Hopefully, it can be assumed that the sign of $P$ and $G$ remember 
the original quantum number~\cite{Banuls:2013jaa}.\footnote{
We check that the $P$ and $C$ give well-defined expectation values 
for the low-lying states of the Schwinger model on the open lattice in the continuum limit.
We also find $|P|,|C|\rightarrow 1$ in the continuum limit 
of the free staggered fermion on the periodic lattice.}
We identify the triplets which have the quantum numbers consistent with 
$J^{PG}=1^{-+}$ and $\bm{J}^2=2$ as the pion states.
From the iso-singlet states with $\bm{J}^2=0$, 
we find the both states of sigma meson $J^{PG}=0^{++}$ and eta meson $J^{PG}=0^{--}$.

After each meson is identified, we obtain the dispersion relation, 
namely the relation between the energy gap $\Delta E_{\ell}=E_{\ell}-E_0$ 
and the momentum square $\Delta K_{\ell}^{2}=\Braket{K^2}_{\ell}-\Braket{K^2}_0$ 
as shown in the right panel of Fig.~\ref{fig:E_and_K2}.
Then the data points are fitted by 
$\Delta E=\sqrt{b^2 \Delta K^2+M^2}$ with fitting parameters $M$ and $b$, 
where the meson mass is given by $M$ as an extrapolation to $\Delta K^2 \rightarrow 0$.
The results are $M_{\pi}=0.426(2)$, $b_{\pi}=1.017(4)$ for the pion; 
and $M_{\sigma}=0.7456(5)$, $b_{\sigma}=1.087(2)$ for the sigma meson with the fitting error.
For the eta meson, we simply solved an equation 
and obtained $M_{\eta}=0.904$ and $b_{\eta}=0.962$.

\section{Conclusion and Discussion}
\label{sec:summary}

In this work, we examine three distinct methods to compute the mass spectrum 
of lattice gauge theories in the Hamiltonian formalism 
for implementation to tensor networks and quantum computation.
We demonstrate the methods using DMRG in the massive 2-flavor Schwinger model at $\theta=0$, 
which shares some properties with $4$d QCD.
There are composite particles (mesons) as triplets or singlets of the isospin symmetry.

\begin{figure}[tb]
\centering
\includegraphics[scale=0.4]{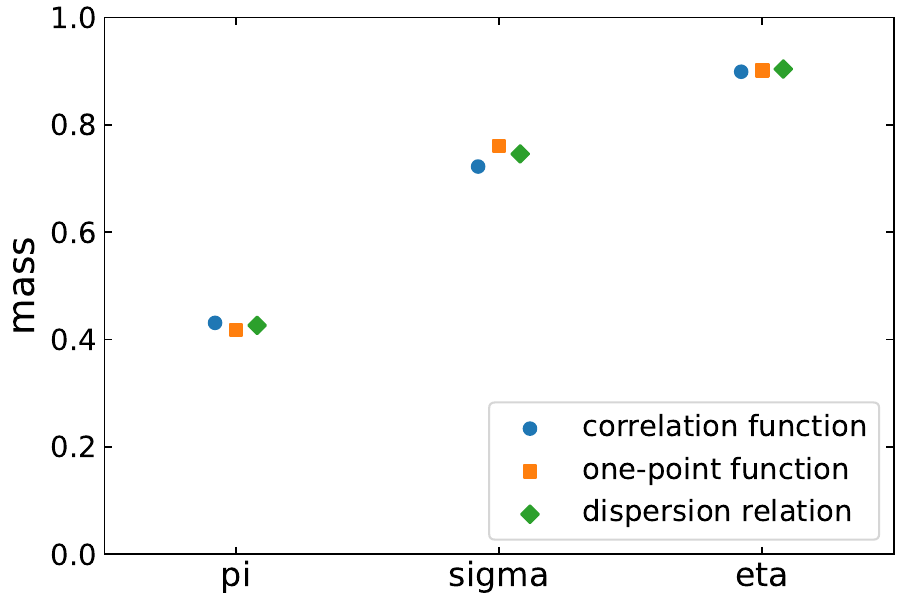}
\vspace{-0.4em}
\caption{\label{fig:mass_comp}
The masses of the pion, sigma, and eta meson computed by the three methods are compared.
Each result is obtained with the given finite lattice spacing.
Although we also put the error bars of the fitting error, they are too small to be seen.}
\end{figure}

We computed the masses of the pion, sigma, and eta meson by the three methods.
The results are shown in Fig.~\ref{fig:mass_comp}.
They are almost consistent with each other up to possible systematic errors, 
such as the continuum and infinite-volume limits.
The results are also consistent with the analytic calculation by the bosonization method.
Indeed, the relation $M_{\pi}<M_{\sigma}<M_{\eta}$ is satisfied, 
and the order of the eta meson mass $M_{\eta}\sim 0.9$ is close to 
$M_{\eta}\sim\mu=g\sqrt{2/\pi}$ since $\mu\sim 0.8$ in the current setup.
We obtained the ratio of the pion and sigma meson mass 
as $M_\sigma/M_\pi=1.68(2),1.821(6),1.75(1)$ 
from the correlation function, the one-point function, 
and the dispersion relation, respectively.\footnote{
Note that the errors in the above values only contain the fitting error, 
and there should be further systematic errors potentially 
coming from the finite lattice spacing, the finite-volume effect, 
the open boundary condition, the cutoff of the bond dimension, etc.}
These results agree with the specific value $M_{\sigma}/M_{\pi}=\sqrt{3}$ 
given by the WKB approximation \cite{Coleman:1976uz} within not more than a 5\% deviation.
The reason why the semiclassical analysis gives such an almost correct answer 
beyond its validity range remains a theoretically interesting question.

So far we have computed the mass spectrum at $\theta = 0$.
It is interesting to extend our work to $\theta \neq 0$, 
where the Monte Carlo simulation suffers from the sign problem \cite{Fukaya:2003ph}.
We expect the three methods to be useful in computing the mass spectrum at $\theta \neq 0$ as well.

\acknowledgments
We would like to thank S.~Aoki, M.~Honda, T.~Nishino, and K.~Okunishi 
for useful discussions.
The numerical calculations were carried out on XC40 at YITP Kyoto U.
and the PC clusters at RIKEN iTHEMS.
The work of A.~M. is supported by FY2022 Incentive Research Projects of RIKEN.
The work of E.~I. is supported by JST PRESTO Grant Number JPMJPR2113, 
JST Grant Number JPMJPF2221, %COI-NEXT
JSPS KAKENHI (S) Grant number 23H05439, %Kiban S
JSPS Grant-in-Aid for Transformative Research Areas (A) JP21H05190, and 
Program for Promoting Researches on the Supercomputer Fugaku” 
(Simulation for basic science: approaching the new quantum era) Grant number JPMXP1020230411.
The work of Y.~T. is supported by JSPS KAKENHI Grant number, 22H01218.
This work is supported by Center for Gravitational Physics and Quantum Information (CGPQI) at YITP.

\bibliographystyle{utphys_short}
\bibliography{./Nf2_Schwinger.bib,./QFT.bib}

\end{document}